\documentclass[twocolumn,showpacs,prep,amsmath,amssymb]{revtex4}

\usepackage{graphicx}
\usepackage{dcolumn}
\usepackage{bm}

\begin{document}
\def\ni{\noindent}
\def\pa{\partial}
\def\vp{v_{\phi}}
\def\Rs{$R_{\odot}$}
\title{Why does the Sun's torsional oscillation begin before the
sunspot cycle?}

\date{\today}
\begin{abstract}
Although the Sun's torsional oscillation is believed to be driven
by the Lorentz force associated with the sunspot cycle, this oscillation
begins 2--3 yr before the sunspot cycle.  We provide a theoretical
explanation of this with the help of a solar dynamo model having
a meridional circulation penetrating slightly below the bottom of the
convection zone, because only in such dynamo models the strong toroidal
field forms a few years before the sunspot cycle and at a higher
latitude.
\end{abstract}

\author{Sagar Chakraborty$^1$}%

\author{Arnab Rai Choudhuri$^{2}$}%
\author{Piyali Chatterjee$^{2}$}%
\affiliation{$^{1}$Department of Theoretical Sciences, S.N. Bose Centre for Basic Sciences, Kolkata - 700098}%
\affiliation{$^{2}$Department of Physics, Indian Institute of Science, Bangalore - 560012}
\maketitle
There is a small periodic variation in the Sun's rotation with the
sunspot cycle, called torsional oscillations.  While this
was first discovered on the Sun's surface [1], the nature
of torsional oscillations inside the solar convection zone was
later determined from helioseismology [2-8].
Several authors [9-12]
developed theoretical models of torsional oscillations by assuming
that they are driven by the Lorentz force of the Sun's cyclically
varying magnetic field associated with the sunspot
cycle.  If this is true, then one would expect
the torsional oscillations to follow the sunspot cycles.
The puzzling fact, however, is that the torsional oscillations of
a cycle begin a couple of years before the sunspots of that
cycle appear and at a latitude higher than where
the first sunspots are subsequently seen.  At first sight, this 
looks like a violation of causality---a classic case of
the effect preceding the cause!   
Our aim is to explain this puzzling
observation, for which no previous theoretical model offered
any explanation. In the models of Covas et al. [10] and
Rempel [12], the theoretical butterfly diagrams extend to
unrealistically high latitudes of about $60^{\circ}$
and the low-latitude branches of
torsional oscillations follow the butterfly
diagrams closely, not starting at higher
latitudes.

Let us summarize some of
the other important characteristics of torsional oscillations,
which a theoretical model should try to explain.  (1) Apart from
the equatorward-propagating branch which moves with the sunspot
belt after the sunspots start appearing, there is also a 
poleward-propagating branch at high latitudes.  (2) The amplitude
of torsional oscillations near the surface is of order 5 m s$^{-1}$.
(3) The torsional
oscillations seem to be present throughout the convection zone,
though they appear more intermittent and less coherent as we go
deeper down into the convection zone (see Figs.\ 4, 5 and 6 in
Howe et al.[7]).  (4) In the equatorward-propagating branch
at low latitudes, the torsional oscillations at the surface seem
to have a phase lag of about 2 yr compared to the oscillations
at the bottom of the convection zone (see Fig.\ 7 in Howe et al.[7]).

The last property of torsional oscillations listed above seems to
suggest that the bottom of the convection zone is the source of the
oscillations, which then propagate upwards. The property (3) then
seems puzzling and contrary to the common sense.  One would expect
the oscillations to be more coherent near the source, becoming
more diffuse as they move upward further away from the source.
The observations indicate the opposite of this.  We shall 
discuss a possible explanation for this observation as well.
Spruit [13] proposed thermal effects near the surface to be
the origin of torsional oscillations---an idea which the property
(4) seems to rule out [7].

While there may not yet be a complete concensus, the majority
of dynamo theorists believe that
the sunspot cycle is produced by a flux transport
dynamo, in which the meridional circulation carries the toroidal
field produced from differential rotation in the tachocline
equatorward and carries the poloidal field produced
by the Babcock--Leighton mechanism at the surface poleward [14-21]. 
Since the differential rotation is stronger at higher latitudes
in the tachocline than at lower latitudes,
the inclusion of solar-like rotation tends to produce a strong
toroidal field at high latitudes rather than the latitudes where
sunspots are seen [17-18]. Nandy \& Choudhuri [19] proposed a hypothesis to
overcome this difficulty.  According to them, 
the meridional circulation penetrates slightly below the bottom
of the convection zone and the strong toroidal field produced at the
high-latitude tachocline is pushed by this circulation
into stable layers below the convection zone where magnetic buoyancy
is suppressed and sunspots are not formed.  
Only when the toroidal field is brought into the 
convection zone by the meridional circulation rising at lower latitudes,
magnetic buoyancy takes over and sunspots finally form.  It may be noted
that there is a controversy at the present time whether the
meridional circulation can penetrate below the convection
zone---arguments having been advanced both against [22] and
for it [23].

If the Nandy--Choudhuri hypothesis (hereafter NC hypothesis) 
is correct, then the toroidal
field of a particular cycle first forms at a relatively high
latitude some time before the sunspots of the cycle would start
appearing.  Assuming that the Lorentz force of the newly formed
toroidal field at the high latitude can initiate the torsional
oscillations, the NC hypothesis provides a natural way to explain
how the torsional oscillations begin at high latitudes before
the appearance of the sunspots of the cycle.  
Our dynamo model based on the NC hypothesis correctly explains 
the onset of torsional oscillations at the high latitude before
the beginning of the sunspot cycle.  We, in fact, would like to
argue that the early onset of torsional oscillations provides a
compelling evidence in support of the NC hypothesis.

Our theoretical model is based on a mean field approach.
However, we know that the magnetic field is highly intermittent
within the convection zone and we need to take account of this
fact when calculating the Lorentz force due to the magnetic field.
Since the convection cells deeper down are expected to have larger
sizes, Choudhuri [24] suggested that the magnetic field within
the convection zone would look as shown in Fig.~1 of that paper.
Demanding that the vertical flux tubes give rise to horizontal
flux tubes with magnetic field $10^5$ G (as suggested by flux
rise simulations [25-28]) after stretching in the tachocline,
the magnetic field inside the vertical flux tubes at the bottom of the
convection zone is estimated to be of order 500 G [24].
This scenario provides a natural explanation for the
properties (3) and (4) of torsional oscillations listed above.
Presumably the torsional oscillation gets initiated in
the lower footpoints of the vertical flux tubes, where the Lorentz
force builds up due to the production of the azimuthal
magnetic field. This perturbation
then propagates upward along the vertical flux tubes
at the Alfven speed.  If the magnetic field inside the flux
tubes is 500 G, then the Alfven speed at the bottom of the
convection zone is of order 315 cm s$^{-1}$ and the 
Alfven travel time from the bottom
to the top turns out to be exactly of
the same order as the phase lag of torsional oscillations
between the bottom of
the convection zone
and the solar surface.  
We admit that the magnetic scenario sketched in Fig.~1 of Choudhuri [24]
and adopted here
is not yet established through rigorous dynamical calculations
and a proper study of the propagation of disturbances through such complex
magnetic structures is unavailable.  However, an assumption
of net upward propagation of magnetic disturbances in spite of all
these complexities is not an unreasonable ansatz, which is justified
by the success of the theoretical model in matching otherwise
unexplained aspects of observational data.
Since the magnetic field at the bottom
is highly intermittent and the velocity
perturbations associated with the torsional oscillations are
likely to be concentrated around the magnetic flux tubes, we
expect the torsional oscillations to be spatially intermittent
at the bottom of the convection zone, as seen in the observational
data [7].  Since the magnetic field near the surface is less intermittent,
the torsional oscillation driven by the Lorentz stress also appears
more coherent there.  We thus have the puzzling situation that
the torsional oscillations seem to become more coherent as they
move further away from the source at the footpoints of flux tubes
at the bottom of the convection zone.

To develop the theoretical model of torsional oscillations, we extend
our already published solar dynamo model [20],
in which the NC hypothesis has been incorporated.  
The basic dynamo code SURYA which is extended for the
present calculations is available upon request.
Apart from the
time evolution equations for the toroidal and poloidal components
of the magnetic field which have to be solved in the dynamo problem,
we also have to solve an additional simultaneous time evolution
equation of the toroidal velocity component $\vp$.  This other equation,
which is essentially the $\phi$ component of the Navier--Stokes equation,
is 
\begin{eqnarray}
\rho \left\{ \frac{\pa \vp}{\pa t} +  \frac{v_r}{r} \frac{\pa}{\pa r} (r \vp) 
+ \frac{v_{\theta}}{r \sin \theta} \frac{\pa}{\pa \theta} (\sin \theta \vp)
\right\} = \nonumber \\ 
({\bf F}_L)_{\phi} +
\frac{1}{r^3} \frac{\pa}{\pa r} \left[ \nu \rho r^4 \frac{\pa}{\pa r} \left( \frac
{\vp}{r} \right) \right] +  \nonumber \\
\frac{1}{r^2 \sin^2 \theta} \frac{\pa}{\pa \theta} \left[ \nu \rho 
\sin^3 \theta \frac{\pa}{\pa \theta} \left( \frac{\vp}{\sin \theta} \right)
\right],
\end{eqnarray} 
where $({\bf F}_L)_{\phi}$ is the $\phi$ component of the Lorentz force.
We use the stress-free boundary condition $\pa \vp/\pa r = 0$ at the solar
surface and take $\vp =0$ at the bottom, although the bottom boundary
condition has no effect when the bottom of the integration region is
taken well below the tachocline as we do.
The kinematic viscosity $\nu$ is primarily due to turbulence within the convection
zone and is expected to be equal to the magnetic diffusivity.  We use the exactly same
profile of $\nu$ as the profile of the diffusivity of poloidal field, which is shown
in Fig.~4 of Chatterjee et al.\ [20].  In other words, we assume the magnetic Prandtl
number to be 1. In order to ensure a period of 11 yr, we choose some parameters
in the dynamo equations slightly different from what were used by Chatterjee
et al.\ [20],
as listed in Table~1 of Choudhuri et el.\ [29].  For the density $\rho$ appearing in (1),
we use the analytical expression used by Choudhuri \& Gilman [25], which gives values of density
consistent with detailed numerical models of the convection zone.

\def\Bp{B_{\phi}}

If the magnetic field is assumed
to have the form
$${\bf B} = B (r, \theta, t){\bf e}_{\phi} + \nabla \times [A(r, \theta, t){\bf e}_{\phi}], \eqno(2) $$
then the Lorentz force is given by the Jacobian
$$4 \pi ({\bf F}_L)_{\phi} = \frac{1}{s^3} J \left( \frac{s B_{\phi}, 
s A }{r, \theta} \right), \eqno(3)$$
where $s= r \sin \theta$.
We, however, have to take some special care in averaging this term, since this is the
primary quadratic term in the basic variables $(A, B, \vp)$ and
has to be averaged differently from all the other linear terms. 
The effect of $\vp$ on the magnetic field is also
quadratic, and has been added to the similar term giving the effect of differential rotation on magnetic fields in the $\phi$-component of the induction equation. 
The $\phi$ component of the Lorentz
force primarily comes from the radial derivative of the magnetic stress $B_r
\Bp /4\pi$ (the term having $B_{\theta} \Bp$ involves $\theta$ derivative
and is smaller). 
This stress arises when $B_r$ is stretched by differential rotation to produce $\Bp$
and should be non-zero only inside the flux tubes.  
We assume that $B_r, \Bp$ are the mean field values, whereas $(B_r)_{\rm ft},  (\Bp)_{\rm ft}$
are the values of these quantities inside flux tubes.  If $f$ is the filling factor,
then we have $B_r = f (B_r)_{\rm ft}$ 
and $\Bp = f (\Bp)_{\rm ft}$, on assuming the same filling factor for both components
for the sake of simplicity. It is easy to see that the mean Lorentz stress is
$$f \frac{(B_r)_{\rm ft}(\Bp)_{\rm ft}}{4 \pi} = \frac{B_r \Bp}{4 \pi f}.$$
This suggests that the correct mean field expression for $({\bf F}_L)_{\phi}$
is given by the expression (3) divided by $f$.  
\begin{figure}
\includegraphics[width=1.15\columnwidth]{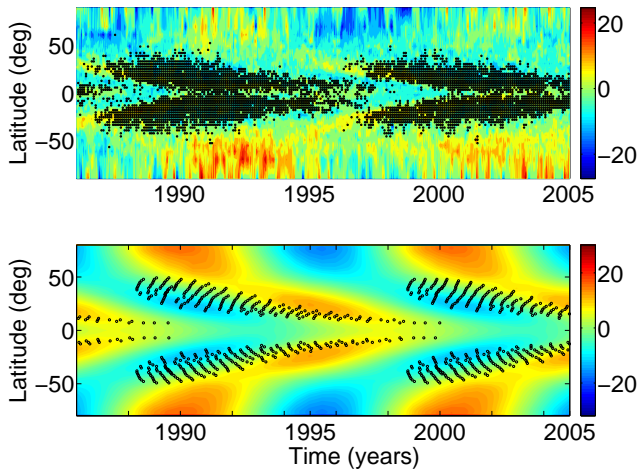}
\caption{{Comparison between observation and theory}.  The upper panel superposes
the butterfly diagram of sunspots on a time-latitude plot of the observed surface zonal
velocity $\vp$ (in m s$^{-1}$) measured at Mount Wilson Observatory (Courtesy: Roger Ulrich).
The comparable theoretical plot is shown in the lower panel, in which the theoretical
butterfly diagram from our dynamo model is superposed on the time-latitude plot
of theoretically computed $\vp$ (in m s$^{-1}$) at the surface. }
\end{figure}

As pointed out by Chatterjee et al. [20], the only nonlinearity in the dynamo equations comes
from the critical magnetic field $B_c$ above which the toroidal field within the
convection zone is supposed to be unstable due to magnetic buoyancy.  Jiang et
al.\ [30]
found that we have to take $B_c = 108$ G (which is the critical value of the mean toroidal
field and not the toroidal field inside flux tubes) to ensure that the poloidal field at
the surface has correct values.  Once the amplitude of the magnetic field gets fixed this
way, we find that only for a particular value of the filling factor $f$ the amplitude of
the torsional oscillations matches observational values. 
 Our calculations give a filling
factor $f\approx 0.067$, which is higher compared to the earlier
estimate of $f \approx 0.02$ by Choudhuri [23].
Theoretical values of velocity in all our figures are computed by using $f =
0.067$. Apart from
the usual meridional circulation used in our model [20], we include
a constant upward velocity 
$v_r =$ 300 cm s$^{-1}$ in (2) to account for the upward transport by
Alfven waves when solving our basic equation (1) for $\vp$. 
Note that this additional $v_r$ does not represent any actual
mass motion and does not have to satisfy the continuity equation
which the meridional circulation satisfies.  Because of our lack
of knowledge about this upward transport, we assume the upward
velocity to be independent of depth and allow it to transport the
magnetic stresses from the bottom to the surface where they freely
move out due to the stress-free boundary condition, mimicking
what we believe must be happening in the real Sun.
 
\begin{figure}
\includegraphics[width=\columnwidth]{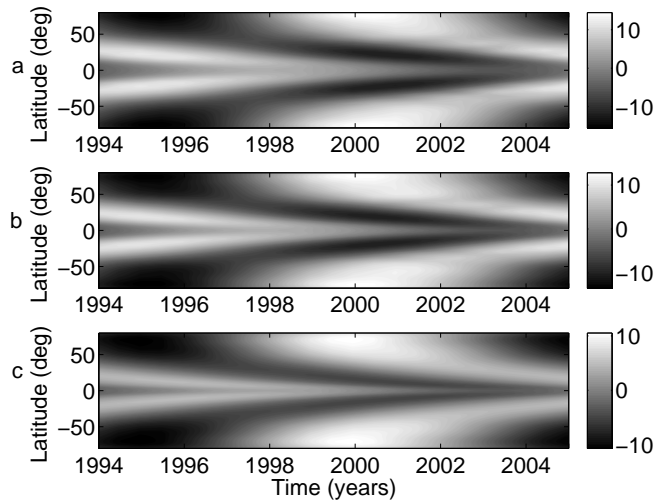}
\caption{{Theoretical torsional oscillations ($\vp$ in m s$^{-1}$)
in time-latitude plots at different depths of the convection
zone: (a) 0.95\Rs, (b) 0.9\Rs, (c) 0.8\Rs}. }
\end{figure}

Figure~1 presents a comparison of theory with observations by putting the butterfly
diagram of sunspots on the time-latitude plot of torsional oscillations at the surface.   The theoretical plot correctly reproduces
the initiation of the low-latitude branch of torsional oscillations about 2 yr
before the sunspot cycle, starting at a latitude higher than typical sunspot latitudes.
Apart from the NC hypothesis, the assumption of the upward advection
of the perturbations at Alfven speed is crucial.  On switching off the Alfven wave,
even though the torsional oscillations begin at a high latitude at the bottom of the
convection zone before the starting of the sunspot cycle, the disturbance has to reach
the surface through diffusion and we do not see the correct initiation of torsional
oscillations at the surface.  We also note that the phase of the torsional oscillations
(i.e. regions of positive and negative $\vp$ in the time-latitude plot)  with respect 
to the sunspot cycle is reproduced quite well.  On decreasing (increasing) the Alfven
speed, the phase of the torsional oscillations with respect to the butterfly diagram gets
shifted towards the right (left). While our main aim was to explain the properties
of the low-latitude branch of torsional oscillations, our theoretical model has
reproduced the high-latitude branch as well, without our having to do anything special
for it. The physics behind this branch will become clear when
we discuss Figure~3 later.
Figure~2 showing torsional oscillations at different
depths has to be compared with the observational 
Figs.~4--6 of Howe et al.\ [7].   A careful look shows a 
small phase delay in the upper layers compared to the lower layers.  
The observational plots
become more intermittent at the greater depths due to the more intermittent nature of
the magnetic field there.  This is not reproduced in the theoretical model based on
mean field equations.

\begin{figure}
\includegraphics[width=\columnwidth]{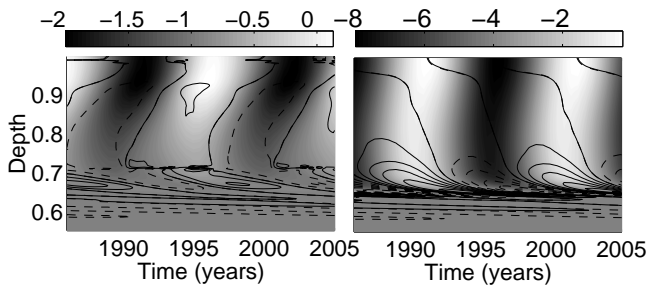}
\caption{{Theoretical torsional oscillations ($\vp$ in m s$^{-1}$) as functions of depth and time at
latitudes $20^{\circ}$ (left) and $70^{\circ}$ (right)}.  
The contours indicate the Lorentz force $({\bf F}_L)_{\phi}$, the solid and
dashed lines denoting positive and negative values. }
\end{figure}

Figure~3 shows how torsional oscillations evolve in depth and time at 2 different
latitudes. The plot for latitude $20^{\circ}$ compares
very well with Fig.~4(D) of Vorontsov et al.\ [5] or 
Fig.~7 of Howe et al.\ [7].  It is clear in the
plot of $20^{\circ}$ that the Lorentz force is concentrated in the tachocline at 0.7\Rs,
where the low-latitude torsional oscillations are launched to propagate upward.
The physics of the high-latitude branch is, however, very different, with the Lorentz
force contours for latitude $70^{\circ}$ 
indicating a downward propagation and not a particularly strong concentration
at the tachocline.  As the poloidal field sinks with the downward meridional circulation
at the high latitudes, the latitudinal shear $d \Omega/ d \theta$ in the convection
zone acts on it to create the toroidal component [21] and thereby the Lorentz stress.
With the downward advection of the poloidal field, the region of Lorentz stress
moves downward.  In the case of the low-latitude branch, the plot for latitude
$20^{\circ}$ shows that the amplitude of the torsional oscillations becomes
larger near the surface due to the perturbations propagating into regions of
lower density, which is consistent with observational
data [5].  If the upward Alfven propagation is switched off, then 
the disturbances from the bottom of the convection zone reach the top by diffusion
(the diffusion time being about 5 yr in our model), but the amplitude of 
torsional oscillations in the upper layers of the convection zone generally falls to
very low values. 

Compared to the earlier theoretical models of torsional oscillations, the
two novel aspects of our model are (1) the NC hypothesis, which allows
the formation of strong toroidal field in the high-latitude tachocline before
the beginning of the sunspot cycle; and (2) the assumption that the perturbations
propagate upward along flux tubes at the Alfven speed.  With these two assumptions
incorporated, our theoretical model readily explains most
aspects of torsional oscillations without requiring any changes in the 
parameters of the original dynamo model [29].  Both these assumptions seem
essential if we want to match theory with observations in detail.

\def\pa{\partial}
\def\st{\sin \theta}
\def\vt{v_{\theta}}
\def\ni{\noindent}

SC thanks C.S.I.R. (India) for financial support through an SRF. 
Partial support from a DST project No.SR/S2/HEP-15/2007 is acknowledged.

\def\apj{{ Astrophys.\ J.}}
\def\aa{{ Astron.\ Astrophys.}}
\def\sol{{ Sol.\ Phys.}}
\def\mnr{{Mon.\ Not.\ R.\ Astron.\ Soc.}}

\end{document}